\documentclass[IEEE,a4paper,12pt]{article}
\usepackage[dvipsnames]{xcolor}
\usepackage[top=1in, bottom=1in, left=1in, right=1in]{geometry}

\usepackage{graphicx} 
\usepackage{tikz}
\usetikzlibrary{positioning}
\usepackage[most]{tcolorbox}
\usepackage{pgfplots}
\pgfplotsset{compat=1.16}
\usepackage{color}
\usepackage{tikz}
\usepackage{fancyhdr}
\usepackage{tikz}
\usepackage{wrapfig}
\usepackage{comment}
\usepackage{algorithm}
\usepackage{algpseudocode}
\usepackage{pgfplotstable}
\algtext*{EndFor}% Remove "end while" text
\algtext*{EndIf}% Remove "end if" text
\newcommand{\nonl}{\renewcommand{\nl}{\let\nl\oldnl}}% 
\algrenewcommand\algorithmicrequire{\textbf{Input:}}
\usepackage{caption}
\usepackage{mathrsfs}
\usepackage{makecell}
\usepackage{diagbox}
\usepackage{siunitx} % Formats the units and values
\usepackage{booktabs}
\usepackage{soul}
\usepackage{multirow}
\usepackage[numbers]{natbib}

\RequirePackage{amsthm,amsmath,amsfonts,amssymb}

\newcommand{\Unif}{\mathrm{Unif}}

\newcommand{\simiid}{\overset{\mathsf{iid}}{\sim}}

\newcommand{\Ncal}{\mathcal{N}}
\newcommand{\Hcal}{\mathcal{H}}

\newcommand{\thr}{{\mathsf{thr}}}

\newcommand{\logl}{\mathsf{lppt}}

\newcommand{\Pa}{\mathrm{P}_a}
\newcommand{\Pb}{\mathrm{P}_b}

\newcommand{\LMmodelnameI}{GPT2~}

\newcommand{\GLMmodelname}{ChatGPT~}

\newcommand{\Pglm}{\mathrm{G}_0}
\newcommand{\Palt}{\mathrm{G}_1}
\newcommand{\Plm}{\mathrm{P}}
\newcommand{\Dkl}{\mathrm{\bar{D}}}

\newcommand{\HC}{\mathrm{HC}^{*}}
\newcommand{\minP}{\min\mathrm{P} }

\usepackage[hidelinks,citecolor=blue]{hyperref}
\hypersetup{colorlinks=true, linkcolor=black}

\newtheorem{prop}{Proposition}

\usetikzlibrary{external,decorations.pathreplacing,calligraphy}
\tikzset{
  prompttext/.style={
    draw,
    rounded corners=5pt,
  align=left, 
  text width=0.45\textwidth,
  execute at begin node=\setlength{\baselineskip}{1em},
  font=\sffamily
  }
}

\tikzset{
  prompthuman/.style={
  prompttext,
  text=blue,
  execute at begin node= {\includegraphics[scale=.1]{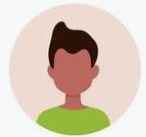}~}
  }
}

\tikzset{
  promptchatgpt/.style={
  prompttext,
  text=black,
  execute at begin node= {\includegraphics[scale=.3]{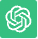}~}
  }
}

\pgfplotstableset{
    highlightrow/.style={
        postproc cell content/.append code={
           \count0=\pgfplotstablerow
            \advance\count0 by1
            \ifnum\count0=#1
            \pgfkeysalso{@cell content=\textbf{##1}}%
            \fi
        },
    },
}

\title{An Information-Theoretic Approach for Detecting Edits in AI-Generated Text}

\author{Idan Kashtan and Alon Kipnis \\ Reichman University}

\begin{document}
\maketitle

\begin{abstract}
    We propose a method to determine whether a given article was written entirely by a generative language model or perhaps contains edits by a different author, possibly a human. Our process involves multiple tests for the origin of individual sentences or other pieces of text and combining these tests using a method that is sensitive to rare alternatives, i.e., non-null effects are few and scattered across the text in unknown locations. Interestingly, this method also identifies pieces of text suspected to contain edits. We demonstrate the effectiveness of the method in detecting edits through extensive evaluations using real data and provide an information-theoretic analysis of the factors affecting its success. In particular, we discuss optimality properties under a theoretical framework for text editing saying that sentences are generated mainly by the language model, except perhaps for a few sentences that might have originated via a different mechanism. Our analysis raises several interesting research questions at the intersection of information theory and data science.
\end{abstract}

\newcommand{\nMonte}{10,000}
\newcommand{\nDocs}{18}
\newcommand{\nDocsD}{11}

\section{Introduction}

\subsection{Background and Motivation}
Suppose an article initially written by a generative language model (GLM) such as ChatGPT (GPT3.5) undergoes relatively minor changes. For example, a human editor adds, removes, or rephrases certain sentences as in the example in Figure~\ref{fig:generation}. This work aims to detect the presence of such edits and, as much as possible, identify edited parts if they exist.
\begin{figure}[H]
    \centering
    \begin{tikzpicture}[thick,scale=1, node distance=1cm]
\node (leftcol) at (0,0) {};
\node (rightcol) at (0.5\textwidth,0) {};
\node[prompthuman, below=0cm of leftcol] (p1) at (0,0) {
     I am interested in writing a Wikipedia-style article about the Welsh Corgi dog breed.};
\node[promptchatgpt, below=.1cm of p1] (p1r) {
    That sounds like a great project! The Pembroke Welsh Corgi is...};
\node[prompthuman, below=.1cm of p1r] (p2) {Can you please write the introduction?};
\node[promptchatgpt, below=.1cm of p2] (p2r) {The Welsh Corgi, often simply referred to as Corgi, is a charming and beloved breed of herding dog with roots deeply entwined in the landscapes and folklore of Wales. Renowned for...};
\node[prompthuman, below=.1cm of p2r] (p3) {Tell me about the history of the Corgi breed.};
\node[promptchatgpt, below=.1cm of p3] (p3r) {
The history of the Corgi breed is deeply rooted in the pastoral landscapes of Wales...
};
\node[prompthuman, below=.1cm of p3r] (p4) {Tell me about ...}; 
% \node[prompttext, below=.5cm of p4] (article) { 
% \vspace{-10pt} \begin{center} Pembroke Welsh Corgi \end{center} 
% Introduction\\
% The Pembroke Welsh Corgi, a small and endearing dog breed known for its distinctive appearance and delightful personality... \\
% \vspace{5pt}
% Appearance \\
% The Pembroke Welsh Corgi boasts a distinctive and charming appearance that...
% };
\node[prompttext, below=0cm of rightcol] (editedarticle) { 
\begin{center}  Welsh Corgi \end{center} 
Introduction\\
The Welsh Corgi, often simply referred to as Corgi, is a charming and beloved breed of herding dog with roots deeply entwined in the landscapes and folklore of Wales. Renowned for their distinctive appearance, characterized by their short legs and elongated bodies, Welsh Corgis have captured the hearts of dog enthusiasts worldwide. 
{\color{red}\st{Beyond their adorable looks}} {\color{blue} Originally bred to herd cattle, sheep, and horses}, Corgis are celebrated for their intelligence, agility, {\color{red}\st{and}}{\color{blue} unwavering loyalty} eagerness to please, {\color{blue}and adorable looks}.
%Original: Beyond their adorable looks, Corgis are celebrated for their intelligence, agility, and unwavering loyalty to their human companions...
%Edited: Originally bred to herd cattle, sheep, and horses, Corgis are celebrated for their intelligence, agility, unwavering loyalty, and eagerness to please.
\\
\vspace{5pt}
History \\
The history of the Corgi breed is deeply rooted in the pastoral landscapes of Wales, where they played a vital role as herding dogs.
{\color{blue} Corgi breeds are classified as Pembroke Welsh Corgi and Cardigan Welsh Corgi, both originating from a common ancestry}.
{\color{red}\st{There are two distinct breeds of Corgis: the Pembroke Welsh Corgi and the Cardigan Welsh Corgi}}.
%Both breeds share a common ancestry but developed separately over centuries.
...
};
\end{tikzpicture}
    \caption{Left: The GLM ChatGPT is sequentially prompted to generate sections of a Wikipedia-style article titled  \texttt{Welsh Corgi}. Right: The composition of the generated text with section titles leads to a so-called GLM-written article. The human editor alters the article in some places. We are interested in detecting the presence of edits if they exist, and their locations.
    }
    \label{fig:generation}
\end{figure}
As can be deduced from Figure~\ref{fig:generation}, we mean ``written by a GLM'' in a relatively broad sense. The pre-edited article may combine a series of GLM outputs produced in response to different human-written prompts or instructions. The situation above might arise when a human editor wishes to improve the GLM text or to hide the fact that the GLM was involved in the writing process altogether.

The popularity of GLM-assisted tools for writing raises interest in detecting text generated by a GLM for several reasons, e.g., to maintain trust and ethical standards in authored material or to study the limitations of GLM technology\footnote{Some companies offering AI content detection tools as of April 2024: GPTZero, Originality, Crossplag, Writer, Copyleaks, OpenAI, Sapling, Content at Scale, Hugging Face, Corrector, Grammar Buddies, Writefull, Hive Moderation, Paraphrasing Tool, QuillBot, Scribrr, Undetectable AI, Stealth Writer, Smodin, Detecting-AI, Crossplag, AI-Purity} \citep{liang2024monitoring, weber2023testing,crothers2023machine,stokel2022ai}. The scenario we address corresponds to a particularly challenging situation in this context because human edits, if present, are typically scattered across the article in locations that are unknown in advance. Furthermore, edits are associated with relatively short pieces of ``text atoms'' like sentences whereas the signal discriminating a GLM from a human in a short text piece may be faint or not exist \citep{zellers2019defending,krishna2023paraphrasing,jakesch2023human,PMID:36635510}. Metaphorically, we seek to detect a few needles in a haystack, uncertain how many needles there are, if any. 
This challenge points in the direction of rare (sparse) and weak signal detection problems in statistics \citep{donoho2004higher,cai2014optimal,mukherjee2015hypothesis,donoho2015special,ke2016detecting,arias2019detection,arias2019detection}. In this work, we propose an approach to detect human edits of mostly GLM text that is based on adapting some of these well-understood tools.

A mirror image of our detection problem is detecting machine text hidden within mostly human text. One interesting motivation for this problem is avoiding undesirable effects of training language models using machine-generated text \citep{oren2023proving,alemohammad2023self, briesch2023large,dohmatob2024model}. We believe that many insights from the analysis in this paper are also relevant to this problem. 

\subsection{Existing Approaches}
Existing approaches for discriminating AI text from human text usually focus on detecting relatively large portions of text \citep{zellers2019defending,krishna2023paraphrasing,crothers2023machine,chakraborty2023possibilities}, whereas we focus on detecting effects that might emerge at sentence level. Due to the apparent rarity of the signal underlying the problem and well-understood limitations of binary detection in the presence of rare and weak features \citep{jin2009impossibility}, it seems that any approach that does not capitalize on the signal's sparsity structure is generally ineffective. For example, machine learning approaches as described in \citep{openAIdetector} and \citep{chakraborty2023possibilities} may be effective if they can learn some lower-dimensional representation of the data. However, such representation does not exist for rare and weak signals with unknown sparsity \citep{jin2009impossibility}. An additional disadvantage of these approaches %with adapting existing machine learning approaches is the challenge of acquiring substantial training data that reflect realistic text edits. Another disadvantage 
is their typical lack of transparency, i.e.,  limited ability to reason a method's outcome and thus limited ability of a user to take actions based on the outcome. 
%. Such reasoning is essential in some applications, e.g., maintaining trust in authored material \citep{gunning2019darpa}.

The problem of detecting edits is also related to the ``style change'' detection problem in mixed authorship studies \citep{bevendorff2022overview,juola2008authorship,stamatatos2009survey,neal2017surveying,zheng2023review}. This problem is considered to be notoriously challenging due to the weakness of the authorship signal when mixed text pieces are short \citep{kestemont2018overview,bevendorff2022overview}. The problem we consider includes the additional difficulty that changes in authorship across an article, if they exist, are few.

\subsection{Our Approach}
Our approach 
%to detecting the presence of human edits within a much larger article written by a GLM relies on well-studied tools from rare and weak signal detection \citep{donoho2004higher,hall2008properties,donoho2015special,ke2016detecting,kipnis2021logchisquared}. It
involves two main steps:
\begin{itemize}
    \item[(i)] Testing the authorship of every sentence individually with respect to the candidate GLM.
    \item[(ii)] Combining the multiple tests to a global test of significance against the null hypothesis of purely GLM text 
    using a method that is sensitive to sparse alternatives.
\end{itemize}
Specifically, we implement Step (i) using the log-perplexity statistic (a.k.a. negative log-likelihood, logloss, crossentropy loss) under a pre-trained large LM that can provide token probabilities. In Step (ii), we use global testing based on Higher Criticism (HC) \citep{donoho2004higher,donoho2015special}.
If a global indication for edits is detected in Step (ii), we add the step:
\begin{itemize}
    \item[(ii')] Report on sentences suspected to be edited. 
\end{itemize}
We implement (ii') using the HC threshold, a by-product of the HC calculation in Step (ii)
\citep{donoho2009feature,donoho2008higher}. The HC threshold is the main reason we use HC rather than other methods that adapt well to sparsity as discussed in  \citep{donoho2004higher,walther2013average,moscovich2016exact}. The entire procedure is illustrated in Figure~\ref{fig:process}. 

\begin{figure}[H]
    \centering
    \begin{tikzpicture}
    \node at (0,0) {
    \begin{tabular}{|p{7cm}|c|c|}
    \toprule
        sentence & $\logl$ & \thead{$\logl$ \\P-value} \\
\midrule
The Welsh Corgi, often simply... & 2.765 & 0.212 \\
Renowned for their distinctive... & 2.864 & 0.2424 \\
\textcolor{blue}{Originally bred to herd cattle,...} & 2.484 & 0.649 \\
This article delves into the... & 3.005 & 0.4904 \\
$\vdots$ & $\vdots$ & $\vdots$\\
Overall, Welsh Corgis have become... & 2.456 & 0.4594 \\
\textcolor{blue}{They have surpassed their humble...} & 3.906 & 0.1057 \\
Their unique combination of historical... & 3.284 & 0.0779 \\
\bottomrule
\end{tabular}};

\draw[line width=1.5pt, dashed] (3.6, 1.5) rectangle (5.2,-2.8);
\draw [decorate, line width=2.5pt, 
    decoration = {brace,
        raise=2pt,
        aspect=0.45}] (5.3,1.5) -- (5.3,-2.8);
\node[right] (HC) at (5.5,-.5) {$\HC$};
\node[above of = HC, xshift=1cm] {$<\mathsf{thr}$};
\node[below of = HC, xshift=1cm] {$>\mathsf{thr}$};

\draw[->] (HC) |- +(1,2) node[right,align=center, line width=1.5pt]  {no \\ edits}; 
\draw[->] (HC) |- +(1,-2) node[right, align=center, line width=1.5pt] {some \\ edits};
\draw[->] (HC)+(1.5,-2.5) -- +(1.5, -3) node[below, align=center, line width=1.5pt] {suspected edit \\ locations};
\end{tikzpicture}
\caption{The detection procedure applied to the example article \texttt{Welsh Corgy} written by the GLM GPT3.5-turbo. Left (table): log-perplexity and its P-value, for each sentence. Actual non-GLM sentences are in blue. Right: The Higher Criticism (HC) score is compared to a threshold, e.g. the $0.95$ quantile of HC under independent and uniformly distributed P-values, or a threshold calibrated via training data. Here log-perplexity is under the language model \LMmodelnameI and the P-values are based on the empirical distribution of sentences from Wikipedia-style articles written by the same GLM. 
    }
    \label{fig:process}
\end{figure}

\subsection{Contributions and Paper Organization}
We describe the method in Section~\ref{sec:description}. In Section~\ref{sec:experiments}, we demonstrate its effectiveness in some realistic scenarios encompassing several text domains. In Section~\ref{sec:IT}, we analyze the method components using tools from information theory, discussing their optimality under a certain theoretical framework that exposes the interesting properties of the problem. Our results promote several open challenges that we discuss in Section~\ref{sec:challenges}.
 
\begin{figure}[H]
    \begin{center}
    \begin{tikzpicture}
    \node[inner sep=0pt] (image1) at (0,0){\includegraphics[scale=.4, trim=1cm 0cm 0.5cm 0cm, clip=false]{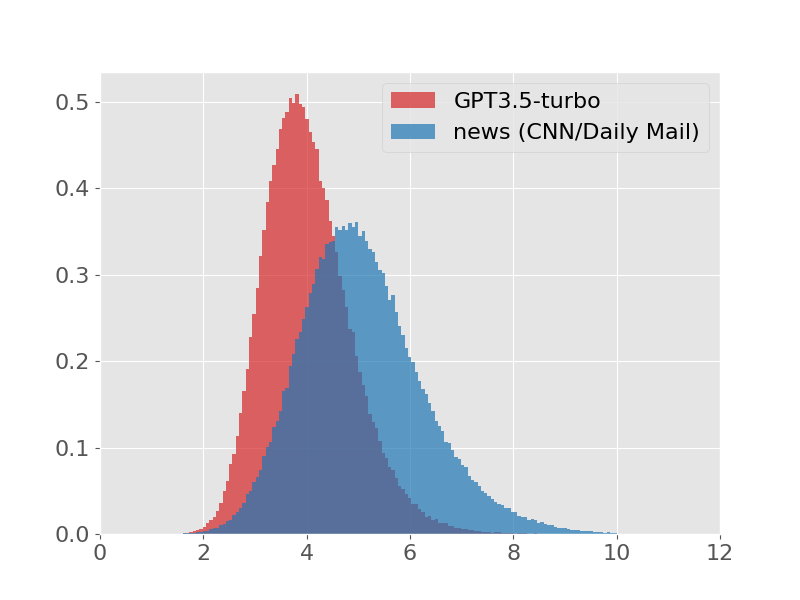}};
    \node[below of=image1, xshift=-0.3cm, yshift=-1.9cm] {$\logl$ [bits/token]~~}; 
    \node[left of=image1, xshift=-2.9cm]{\rotatebox{90}{frequency}};

    \node[inner sep=0pt] (image2) at (7.5,0){\includegraphics[scale=.4, trim=5.5cm 0cm 3cm 0cm, clip=true]{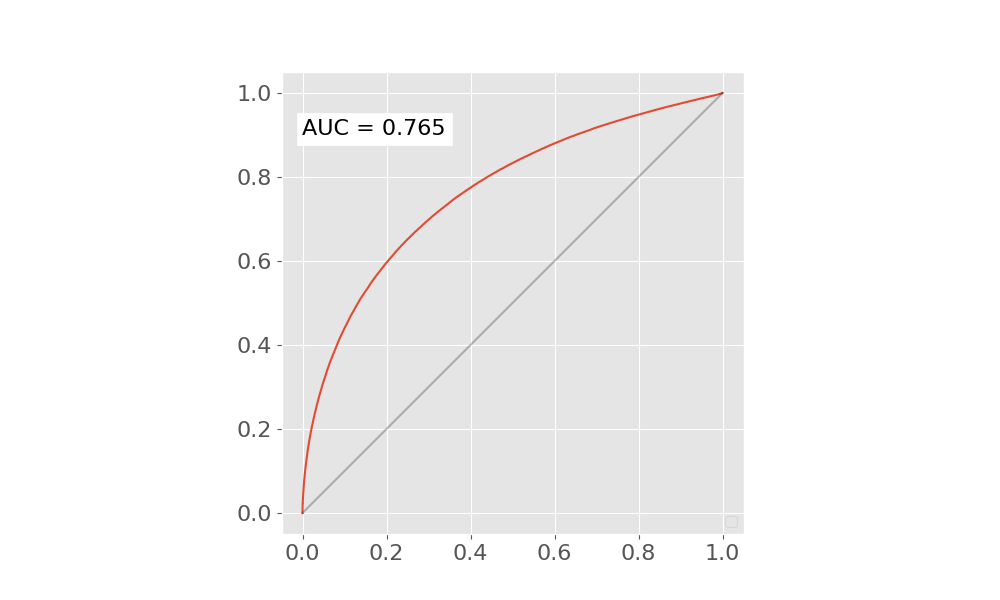}};
    \node[left of = image2, xshift=-2.5cm] {\rotatebox{90}{true positive rate}};
    \node[below of = image2, xshift=-0.5cm, yshift=-2cm] {{false positive rate}};
    \end{tikzpicture}

    \begin{tikzpicture}
    \node[inner sep=0pt] (image1) at (0,0){\includegraphics[scale=.4, trim=1cm 0cm 0.5cm 0cm, clip=false]{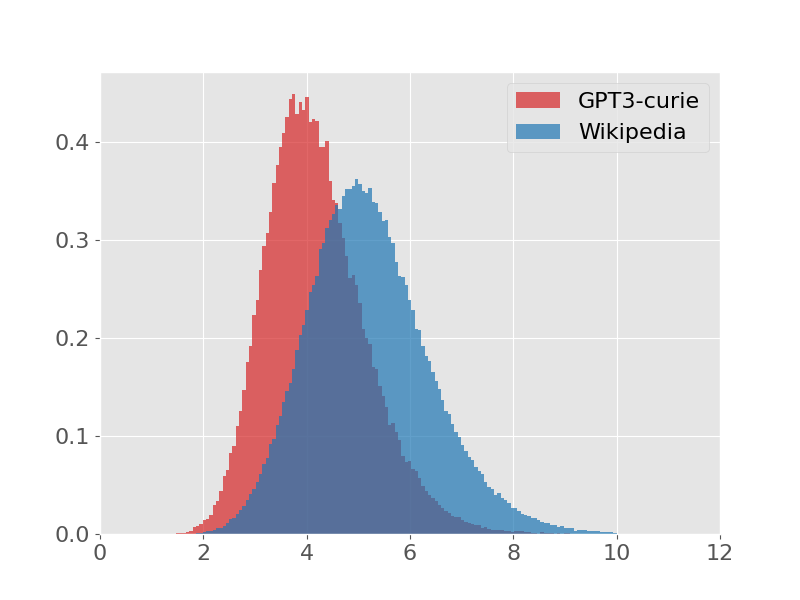}};
    \node[below of=image1, xshift=-0.3cm, yshift=-1.9cm] {$\logl$ [bits/token]~~}; 
    \node[left of=image1, xshift=-2.9cm]{\rotatebox{90}{frequency}};

    \node[inner sep=0pt] (image2) at (7.5,0){\includegraphics[scale=.4, trim=5.5cm 0cm 3cm 0cm, clip=true]{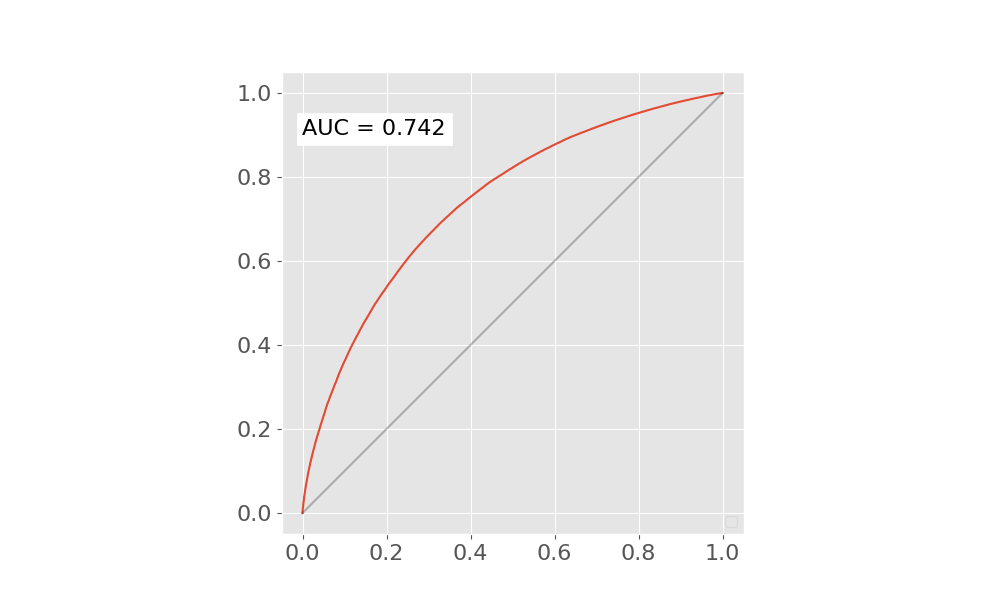}};
    \node[left of = image2, xshift=-2.5cm] {\rotatebox{90}{true positive rate}};
    \node[below of = image2, xshift=-0.5cm, yshift=-2cm] {{false positive rate}};
    \end{tikzpicture}

\caption{Discriminating GLM from non-GLM sentences using the log-perplexity (LPPT) statistic \eqref{eq:LL_def}. Left: histogram by class of LPPT of sentences from the dataset 
    \texttt{News Articles} \citep{isarth_2023} (top) and \texttt{Wikipedia Introductions} 
    %\texttt{Scientific Abstracts} (middle) \citep{sivesind_2023},
    \citep{aaditya_bhat_2023} (bottom). Right: the receiver operating characteristic (ROC) of a test based on the LPPT. The area under the ROC curve (AUC) is indicated. In both cases, LPPT is under the language model GPT2 (1.5B). %\note{Add Evaluations using \texttt{Phi2}.
    }
    \label{fig:individual_chunks}
\end{center}
\end{figure}

\section{Method Description}
\label{sec:description}

In this description and throughout the paper, it is useful to distinguish between two types of language models based on the output they provide. 
\begin{itemize}
\item (predictive) Language Model (LM): provides a probability distribution over a dictionary of tokens conditioned on an input context. We typically denote such a model by $\Plm$. 
    \item Generative Language Model (GLM): produces sequences of tokens in response to an input context. We typically denote such a model by $\Pglm$.  
\end{itemize}
Any GLM can be conceptually treated as a LM since it induces a distribution over tokens; we distinguish the two types because in many situations the next token probabilities of the GLM are inaccessible to the user, e.g. the ChatGPT interface \citep{chatgpt2022} does not provide such probabilities as opposed to the open source LMs Llama \citep{touvron2023llama} and Falcon \citep{falcon40b}.

\subsection{Testing individual sentences}
We think about a sentence $S$ as a sequence $t_{1:|S|} = (t_1,\ldots,t_{|S|})$ of tokens from a finite dictionary. Given a LM $\Plm$, the (normalized) log-perplexity (LPPT) of $S$ with respect to $\Plm$ is 
\begin{align}
    \label{eq:LL_def}
\logl(S;\Plm) := -\frac{1}{|S|} \sum_{i=1}^{|S|} \log \Plm(t_i | t_{1:i-1}) 
\end{align}
(in \eqref{eq:LL_def} and throughout we use the notation $t_{1:1}=\emptyset$.). Other names for the LPPT statistic are the negative log-likelihood under $\Plm$, the logarithmic loss with respect to $\Plm$ \citep{bishop2006pattern}, the self-information of $\Plm$ evaluated at $S$ \citep{merhav1998universal}, and crossentropy with a mass distribution at the sequence $S$. We prefer the name LPPT because of the association of perplexity with natural language processing \citep{jurafsky2023}. Throughout this paper, we use 2 as the basis of the logarithm, hence $\logl(S;\Plm)$ is measured in bits per token.

%Several previous works have used the LPPT statistic to distinguish GLM from non-GLM texts \citep{zellers2019defending,gehrmann2019gltr,pu2023deepfake}; it is likely used in many commercial detectors.

An empirical result illustrated in Figure~\ref{fig:individual_chunks} says that under a specific $\Plm$, sentences written by a particular GLM tend to have lower values of $\logl(S;\Plm)$ than sentences written by humans in a similar context. This observation justifies using the LPPT statistic to test the authorship of individual sentences, i.e. testing against the null
\begin{align}
    \label{eq:null_sentence}
H_{0}(S) \, :\, \text{``sentence $S$ was written by the GLM $\Pglm$.''}
\end{align}
The right-hand side of Figure~\ref{fig:individual_chunks} illustrates the receiver operating characteristic (ROC) curves of the LPPT test against the null \eqref{eq:null_sentence} under different datasets. Given a document partitioned into sentences $D=(S_1,\ldots,S_n)$, we summarize the evidence against $H_0(S_i)$ using a P-value:
\begin{align}
    \label{eq:P-value}
p_i := p(S_i) := \Pr_{S \sim \Pglm}\left[S \geq \logl(S_i; \Plm)\right].
\end{align}
The evaluation of $p_i$ requires the distribution of $\logl(S;\Plm)$ for $S \sim G_0$, represented by the red histograms in Figure~\ref{fig:individual_chunks}. As it turns out, this distribution is affected by several factors including the text's domain and the length of every sentence as we discuss in Section~\ref{sec:refinements}. We adjust this distribution for sentence length in all P-value evaluations in this paper.
The table in Figure~\ref{fig:process} shows examples of LPPT and the corresponding length-adjusted P-values of several sentences from the example in Figure~\ref{fig:generation}. s

We note that the non-trivial power of the classifier based on LPPT values observed in each case in Figure~\ref{fig:individual_chunks} suggests that, for long enough documents, it is possible to reliably separate between the class of documents written entirely by the GLM and the class of non-GLM documents. Indeed, fix one of the dataset cases in Figure~\ref{fig:individual_chunks} and consider a simple model in which a document is generated by independently sampling sentences from one of the distributions represented by the histograms of this case. Use the likelihood ratio test of the LPPTs of individual sentences to classify a document to GLM or non-GLM. The error probability of this test can be made to vanish rapidly as the number of sentences increases (e.g., this follows from the Chernoff–Stein Lemma \citep{cover2006elements}.). This note emphasizes that the problem this paper considers -- separating the class of documents written by the GLM from the class of documents that contain mostly GLM-written text with some non-GLM edits -- is much more challenging.  

\subsection{Global Testing using Higher Criticism (HC)}

\begin{figure}[H]
    \centering
    \begin{tikzpicture}
        \node (image) at (0,0) {
    \includegraphics[scale=.5]{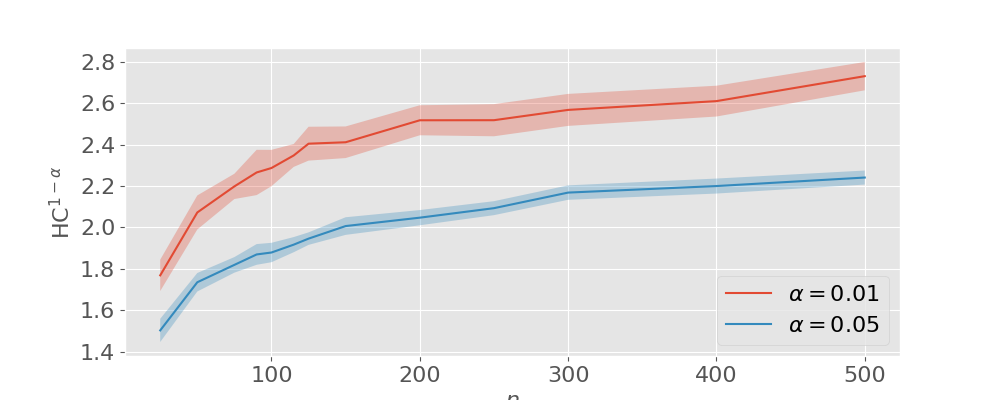}    
        };
    \node[fill=white,left of = image, xshift=-4.7cm, yshift=0cm] {\rotatebox{90}{$\HC_{1-\alpha}$}};
    \node[fill=white,below of = image, xshift=0cm, yshift=-1.6cm] {$n$ [sentences]};
    \end{tikzpicture}
    \caption{
    Simulated critical values for a test of significance level $\alpha$ based on Higher Criticism of $n$ independent P-values. The number of samples in each configuration is $\nMonte$. Bootstrapped $0.95$ confidence intervals are indicated.}
    \label{fig:critical_values}
\end{figure}

We combine the per-sentence P-values $p_1,\ldots, p_n$, $n=|D|$, of \eqref{eq:P-value} to a single value using HC \citep{donoho2004higher,donoho2008higher,donoho2015special}:
\begin{align}
    \label{eq:HC_def}
        \HC := \HC(p_1,\ldots,p_n) := \max_{1 \leq j \leq n \gamma_0} \mathrm{HC}_j,\qquad \mathrm{HC}_j := \sqrt{n}\frac{\frac{j}{n} - p_{(j)}}{ \sqrt{\frac{j}{n} \left( 1- \frac{j}{n} \right)}}.
\end{align}
Here $p_{(1)} \leq \ldots \leq p_{(n)}$ are the order statistics of the P-values and $\gamma_0 \in (0,1)$ is a fixed parameter that limits the range of P-values involved. Typically, $\gamma_0$ does not affect the large sample behavior of $\HC$ and is taken to be in the range $(0.1,0.5)$ \citep{donoho2015special}; we use $\gamma_0 = 0.25$ in this paper as this choice appears to provide good results. HC is known to be sensitive to departures in a small and unknown set of individual tests, thus it is useful as an index of discrepancy between the two classes, indicating that the document was edited for large values of $\HC$. This property leads to a binary classifier whose threshold ($\mathsf{thr}$ in Figure~\ref{fig:process}) can be calibrated, e.g., by a held-out dataset. We can also use HC as level $\alpha$-test against the global null
\begin{align}
    \label{eq:global_null}
H_0 = \bigcap_{S \in D} H_0(S) = \text{``The document was written entirely by $\Pglm$''},
\end{align}
by setting $\mathsf{thr} = \HC_{1-\alpha}$, where $\HC_{1-\alpha}$ is the $1-\alpha$ quantile of $\HC$ under $H_0$ for $\alpha \in (0,1)$, e.g. $\alpha=0.05$. We may estimate $\HC_{1-\alpha}$ from the data using documents from the null class, i.e., written entirely by $\Pglm$, if these are sufficiently available. Otherwise, we may simulate critical values under $H_0$ provided some conditions are met. Specifically, when the P-values are independent and uniformly distributed under $H_0$, the asymptotic distribution of $\HC$ under $H_0$ as $n\to \infty$ follows that of a maximum Brownian bridge, although it may be significantly stochastically smaller in finite samples \citep{donoho2004higher}. For this reason, it is common to simulate critical values for a test based on $\HC$ for specific sample sizes as illustrated in Figure~\ref{fig:critical_values}. In practice, LPPT values of sentences are likely to be dependent since the sentences are. The critical values of $\HC$ are known to be relatively unaffected when the P-values experience a form of short-term dependency \citep{delaigle2009higher}. Under other types of dependency, the test may experience a reduction in power \citep{hall2008properties, hall2010innovated}. 
%unless the HC statistic is properly calibrated \citep{hall2010innovated}. 
For this reason, if possible, we recommend estimating $\HC_{1-\alpha}$ based on complete documents from the null class to improve the test's power. 

\subsection{Identifying edited sentences
\label{sec:HCT}
}
When the HC test rejects $H_0$, the set \begin{align}
    \label{eq:HCT}
    I^* := \{i\,:\, p_i \leq p_{j^*}\},\quad 
    j^* := \arg \max_{1 \leq j \leq n \gamma_0} \HC_j,
\end{align}
corresponds to P-values affecting $\HC$ most and thus providing the strongest evidence against $H_0$. This set is known to have interesting optimality properties in the context of feature selection for binary classification \citep{donoho2009feature,donoho2008higher}. We use this set to indicate sentences that we suspect are not written by the GLM; we may want to examine the authorship of these sentences manually or using other means that do not necessarily rely on the LPPT statistic. 

We summarize the full procedure in Algorithm~\ref{alg:1}, and illustrate it using an example text in Figure~\ref{fig:process}. Table~\ref{tbl:HCT} shows sentences included in $I^*$ from the article generated in the example in Figure~\ref{fig:generation}. 

\begin{algorithm}[ht]
\begin{algorithmic}
\Require{ language model $\Plm$; 
document $D=(S_1,\ldots,S_n)$; 
survival function $\bar{F}_{\Pglm;\Plm}$
of the LPPT of sentences from $\Pglm$ under $\Plm$; threshold $\mathrm{thr}$ (e.g, $\mathrm{thr}=\HC_{1-\alpha}$)}
\State {\color{gray} \# Step I: Testing individual sentences:}
\For {$S_i \in D = (S_1,\ldots,S_n)$}
\State $l_i \leftarrow \logl(S_i;\Plm )$ 
\State $p_i \leftarrow \bar{F}_{\Pglm;\Plm}(l_i)$ 
\EndFor
\State {\color{gray}\#  Step II: Global testing using HC:}
\If {$\HC(p_1,\ldots,p_n) > \mathrm{thr}$,} 
\State {\text{reject $H_0$}}
\State {\color{gray}\#  Step II': Report suspected edits:}
\State {return $\{S_i,\,:\,p_i \leq p_{i^*} \}$} as suspected edits 
\Else 
\State {\text{do not reject $H_0$}}
\EndIf
\end{algorithmic}
\caption{Test whether a document $D$ was written by the language model $\Pglm$ or not
\label{alg:1} }
\end{algorithm}

\begin{table}[ht]
    \centering
    \begin{tabular}{|p{10cm}|c|}
    \toprule
        sentences in $I^*$ & \thead{LPPT \\ P-value} \\
\midrule
Despite their herding heritage gradually diminishing,... & 0.0113 \\
Corgi-themed fundraisers and charity events have... & 0.0211 \\
Legend has it that the fairies... & 0.0346 \\
\textcolor{blue}{It is believed that the Cardigan...} & 0.0400 \\
From their origin as indispensable herding... & 0.0417 \\
Cardigan Corgis were also adept herding... & 0.0435 \\
Their unique combination of historical significance,... & 0.0779 \\
\textcolor{blue}{They have appeared in several animated...} & 0.0820 \\
\textcolor{blue}{They have surpassed their humble origins...} & 0.1057 \\
$\vdots$ & $\vdots$ \\
Mascots and Symbols: In some regions,... & 0.1883 \\
Here are some ways in which... & 0.1942 \\
\textcolor{blue}{The breeds are named for the...} & 0.2065 \\
\textcolor{blue}{A Welsh Corgi appeared with Queen...} & 0.2114 \\
The Welsh Corgi, often simply referred... & 0.2120 \\
\bottomrule
\end{tabular}
    \caption{Sentences from the article \texttt{Welsh Corgi} of Figures~\ref{fig:generation} and \ref{fig:process} that the HC threshold identified as potential non-GLM sentences (edits). Actual non-ChatGPT sentences are in blue. We emphasize that the inclusion of a sentence in this set is based on the global procedure in \eqref{eq:HCT} rather than on the individual significance of the sentence's P-value.
    }
    \label{tbl:HCT}
\end{table}

\begin{figure}[H]
    \centering
    \begin{tikzpicture}
    \node (image) at (0,0) {\includegraphics[scale=.35]{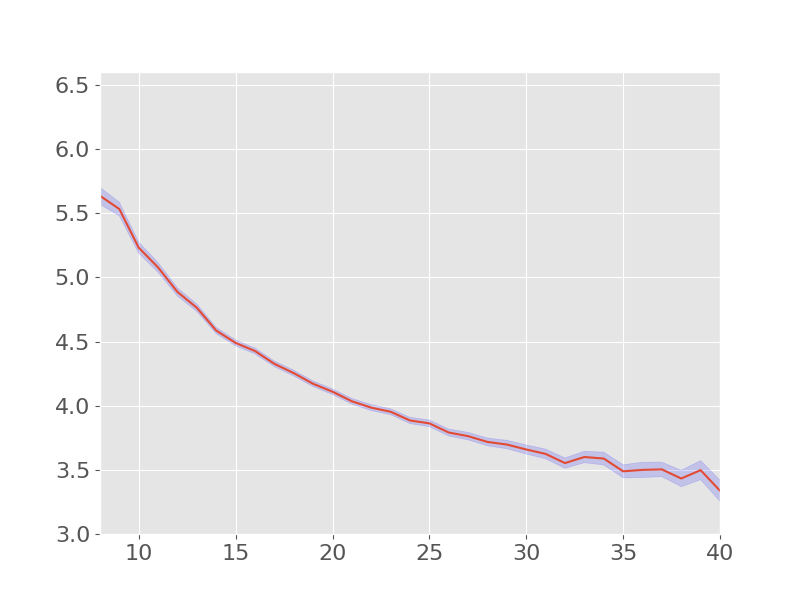}};
    \node[left of = image, xshift=-2.4cm, yshift=0cm] {\rotatebox{90}{$\logl$ [bits/token]}};
    \node[below of=image, xshift=0, yshift=-1.5cm] {length [tokens]};
    \end{tikzpicture}
    \begin{tikzpicture}
    \node (image) at (0,0) {\includegraphics[scale=.35]{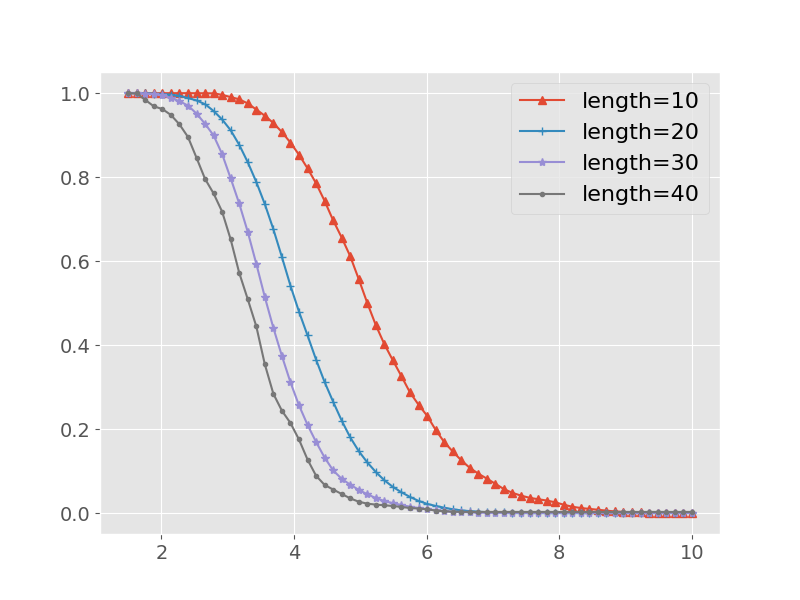}};
    \node[left of = image, xshift=-2.4cm, yshift=0cm] {\rotatebox{90}{probability}};
    \node[below of=image, xshift=0, yshift=-1.5cm] {$\logl$ [bits/token]};
    \end{tikzpicture}
%{figs/logperp_vs_len_first_sentence_gpt2-xl_no_context_wiki-long.png}
    
    \caption{Adjusting the perplexity test for the number of tokens in a sentence. Left: averaged log-perplexity versus sentence length. 
The shaded area indicates 2 standard errors. Right: fitted log-perplexity survival functions of GPT2 for several lengths. Based on 20,000 samples from the dataset \texttt{Wikipedia Introductions} \citep{aaditya_bhat_2023}.
    }
    \label{fig:refinements}
\end{figure}

\subsection{Refinements and Generalizations\label{sec:refinements}}

\subsubsection{Adjusting the log-perplexity distribution for sentence's length}
Tokens appearing later in the sentence tend to be more reliably predicted than tokens at the beginning, 
a phenomenon observed in \citep{shannon1951prediction}. It follows that the average log-perplexity tends to be smaller for longer sentences, as illustrated in Figure~\ref{fig:refinements}. We can therefore attain better sensitivity of the perplexity test by comparing the LPPT of the $i$-th sentence $S_i$ to the distribution of LPPT of sentences produced by $\Pglm$ with the same length as $S_i$. Formally, this means replacing the test \eqref{eq:P-value} with
\begin{align}
    \tilde{p}_i := \Pr_{S \sim \Pglm,} \left[S \geq \logl(S_i; \Plm) \mid |S|=|S_i|\right], 
    \label{eq:P-value_w_length}
\end{align}
and thus the survival function $\bar{F}_{\Plm,\Pglm}$ in Algorithm~\ref{alg:1} receives two parameters: the LPPT of $S_i$ and its number of tokens $|S_i|$. In practice, we estimate $\bar{F}_{\Plm,\Pglm}$ for every possible number of tokens. When the number of data points for calibration is somewhat scarce, a curve-fitting estimate is useful since $\bar{F}_{\Plm,\Pglm}$ appears to vary smoothly with the number of tokens. 

Another factor affecting the perplexity is the sentence's location within the document. For example, the first sentence in every paragraph appears to have higher perplexity than subsequent sentences. We leave the adjustment of our method to this factor as future work. 

\subsubsection{Unusually short and long sentences}
Our experience shows that the perplexity detector is ineffective for sentences of about $10$ tokens or fewer. We excluded such sentences from the process and did not evaluate their P-values. 

We found it difficult to estimate the LPPT distribution of sentences of more than 50 tokens since they are very infrequent in our data. We only consider the first 50 tokens when evaluating the LPPT of such sentences.

\subsubsection{Generalizing Step I: Testing pieces of text individually}
Our method uses sentences as text atoms and considers their LPPT. Natural generalizations of this step include the considerations of other pieces of text like paragraphs, as well as detectors that are not necessarily based on the perplexity, e.g., probability curvature \citep{mitchell2023detectgpt}, word-frequencies \citep{mosteller2012applied}, and other feature \citep{mindner2023classification}.

\subsubsection{Generalizing Step II: Inference based on multiple testing}
HC is just one approach for testing the global significance of individual tests, motivated by the rare editing model over sentences and the sensitivity of HC to rare effects. Under deviations from this model or due to other considerations, 
other methods from multiple comparisons and meta-analyses in statistics may be preferable \citep{rupert2012simultaneous,benjamini2010simultaneous,efron2012large}. For example, instead of HC, we may combine P-values using Fisher's method
\begin{align}
    \label{eq:Fisher}
    F_n := F_n(p_1,\ldots,p_n):=-2\sum_{i=1}^n\log(p_i).
\end{align}
$F_n$ is known to be effective in detecting many relatively frequent but potentially very faint effects \citep{arias2011global,kipnis2021logchisquared}. Therefore, $F_n$ can be used when we test $H_0$ of \eqref{eq:global_null} against an alternative specifying that the GLM text has gone through substantial editing.

Another alternative to inference based on HC is useful when we are interested in selecting a set of suspected edits with some control over the proportion of falsely reported edits. In this case, the Benjamini-Hochberg (BH) false discovery rate (FDR) controlling procedure to the P-values in \eqref{eq:P-value} may be useful  \citep{benjamini1995controlling}. We note that the BH procedure is in general less powerful for global testing than HC. Namely, it is possible that while HC correctly finds the body of P-values significant, the BH procedure with an FDR parameter $\alpha$ may report on an empty set of P-values with probability at least $1-\alpha$, for every $\alpha\in(0,1)$ \citep{jin2003detecting,kipnis2021logchisquared}.

\section{Empirical Results
\label{sec:experiments}}

We conducted extensive simulations using publicly available datasets and new datasets that we created. The new datasets and the code for obtaining all the results are available in the link at the end of this paper. 

We tried several publicly available LMs for the detection model $\Plm$ in Algorithm~\ref{alg:1}, including GPT2 1.5B parameters \citep{radford2019language}, Falcon 7B parameters \citep{falcon40b}, Llama 7B parameters \citep{touvron2023llama}, and Phi2 2.7B parameters \citep{javaheripi2023phi}. We only report on results with GPT2 1.5B parameters (aka. GPT2xl) and Phi2 since these models attained the highest area under the ROC curve in the binary detection problem of individual sentences for all datasets we considered. We discuss in Section~\ref{sec:challenges} the open challenge of selecting or crafting $\Plm$ with optimal detection properties.

We experimented with data created by the GLMs GPT3-curie and GPT3.5-turbo (ChatGPT) arranged in 5 datasets as we explain in detail below. We tried to generate data using publicly available GLMs not in the GPT family, but they did not produce articles of satisfactory quality. 

In the sections below we report the method's performance under different settings and data. 

\iffalse
\begin{table}[h!]
    \centering
    \begin{tabular}{|l|c|c|c|c|}
    \hline
    & GPT2 & Phi2 & Falcon(7B) & Llama2(7B) \\
    \hline 
        \texttt{wiki} & 0.775 & 0.791 & 0.710 & 0.754\\ 
        \hline        
         \texttt{news} & 0.734 & 0.781 & 0.655 &  0.654 \\
         \hline
         \texttt{abstracts} & 0.771 & 0.81 & 0.680 & 0.722 \\
         \hline
    \end{tabular}
    \caption{Language models and their ability to distinguish GLM from non-GLM sentences using the perplexity statistic. The table indicates the area under the receiver operating characteristic curve (AUC-ROC) in binary classification based on the log-perplexity statistic \eqref{eq:LL_def}, for the three datasets described in Section~\ref{sec:simulated_edits}}. 
    \label{tbl:separation}
\end{table}
\fi

\subsection{
Power analysis using mixed machine and human sentences
\label{sec:simulated_edits}}

We first demonstrate the method's effectiveness using a synthetic dataset of articles of mixed authorship involving GLM and non-GLM text. We generated each article by sampling sentences from the non-GLM article and inserting those into the GLM article at random locations. Since the GLM and non-GLM articles are on the same topic, the mixed article is typically coherent in content hence the situation simulates well a GLM text edited in a few locations. As raw data for mixing, we use the three datasets listed below, in which every entry has two articles under the same title, one written entirely by a GLM and one written by a human or several humans. 
\begin{itemize}
    \item \texttt{Wikipedia Introductions} \citep{aaditya_bhat_2023}. Each entry corresponds to a Wikipedia article. The dataset contains the several first sentences of the Introduction of this article as non-GLM text and text generated by GPT3-curie in response to a relevant prompt as a GLM-written article. We excluded entries from this dataset in which the length of the GLM text was less than $15$ sentences, resulting in a total of $9,821$ remaining entries.
    \item \texttt{News Articles} \citep{isarth_2023}. Each entry contains a news article, its highlights as provided by a human annotator, and an article generated by GPT3.5-turbo from these highlights. The dataset has $20,000$ entries. 
    %We removed from this dataset the entries in which the GLM text is shorter than 15 sentences.
    %We removed articles containing less than 10 sentences. 
    \item \texttt{Scientific Abstracts}  \citep{sivesind_2023}. Each entry contains the abstract of a scientific research paper and text produced by GPT3.5-turbo in response to a prompt requesting a paragraph of text with similar properties. The dataset has $10,000$ entries.
\end{itemize}
We divided all entries within each dataset into roughly 10 equally sized groups. For a given edit ratio and article length, we report on results averaged over $10$ iterations in a cross-validation fashion: In Iteration $i$, we simulated edits only in Group $i$, leaving the other groups unaffected and using them to characterize the null LPPT distribution $G_0$ and estimate $\HC_{1-0.05}$. We arranged entries in either group randomly and consolidated them into articles according to the prescribed number of sentences, truncating excess sentences. Each article in Group $i$ is then modified by inserting sentences randomly sampled from the corresponding non-GLM articles according to the mixing ratio $\epsilon$. We set $\thr_i$, our estimate of $\HC_{1-0.05}$, based on the remaining articles: $\thr_i$ is the $0.95$-th quantile of $\HC$ value of articles not in Group $i$. We estimate the power by the fraction of articles in Group $i$ exceeding $\thr_i$. We repeat this process for $i=1,...,10$ and report on the average across all groups as the power estimate. 

\begin{figure}[H]
\begin{center}
\includegraphics[scale=0.4]
{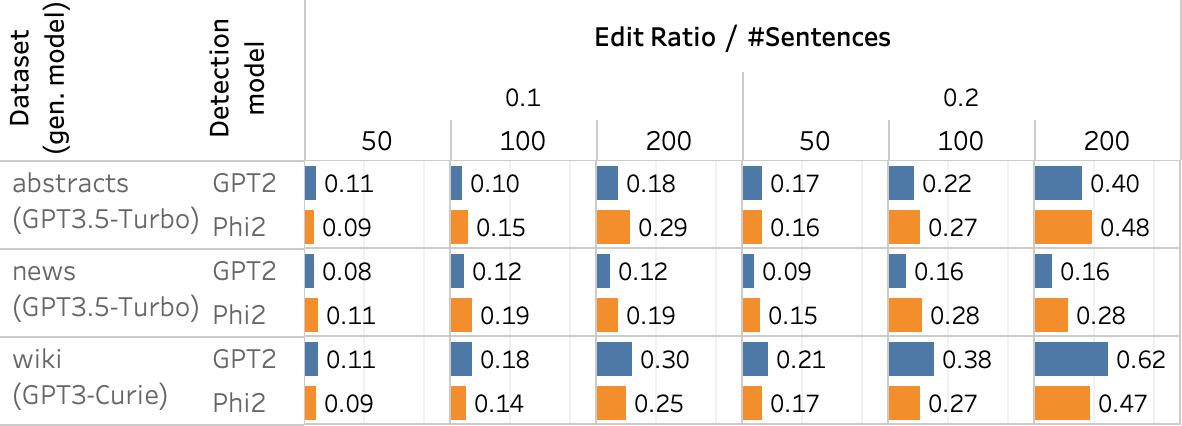}
\end{center}
    \caption{Estimated 
    power at level $\alpha=0.05$ in detecting simulated edits versus the number of sentences in an article and the fraction of edited sentences. We compared two detection models (GPT2 and Phi2) across $3$ datasets. The estimated power is the true positive rate averaged over 10 splits; the standard error in all cases is less than 0.01. The power generally increases with the fraction of non-GLM sentences and article length. } 
    \label{tbl:results}
\end{figure}

The resulting power estimates are shown in Figure~\ref{tbl:results}. It appears that our method has non-trivial power for an editing rate as low as $10\%$ of the sentences, for articles as short as $50$ sentences. The power generally increases with the editing rate and the length of the text and varies between datasets and the two detection models. 

\subsection{Realistic edits of Wikipedia-style articles by topic 
\label{sec:realistic_edits}
}
We created a dataset of Wikipedia-style articles using GPT3.5-turbo by repeatedly prompting this GLM to write article sections\footnote{Via OpenAI's API.}, similarly to Figure~\ref{fig:generation}. We used titles of articles from Wikipedia falling into 5 topics, roughly 200 articles per topic. Article titles were randomly selected within the topic, provided they satisfy our inclusion criteria: at least 5 sections within the article. Section titles in the prompts for generating each article are taken from the corresponding Wikipedia article. The number of sentences in each article is between 50-300 with an average of about $180$.

We simulated realistic edits in this dataset by randomly sampling sentences from the actual Wikipedia article and inserting each under the same section name. Therefore, the resulting article has a realistic structure. Our experience shows that it is very challenging for a human to reliably determine which sentences, if any, were not written by the GLM. 

\begin{figure}[H]
    \centering
    \includegraphics[scale=0.4]{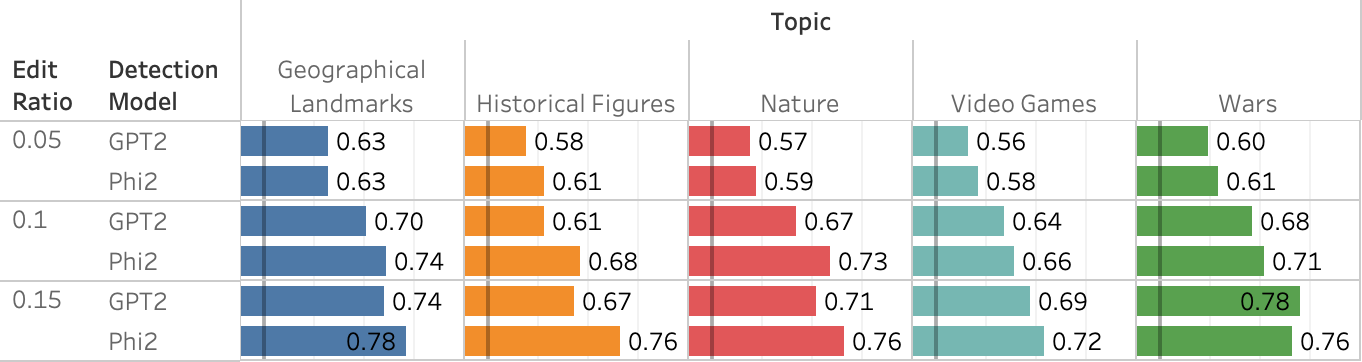}
    \caption{Accuracy in detecting the presence of human text within GPT3.5-turbo articles, by topic. We compare the accuracy across $5$ topics, $3$ edits ratios, and $2$ detection models (GPT2 and Phi2). 
    %Each article is created by prompting based on a Wikipedia page and edited by mixing actual sentences from that page into the GPT4-written article. Every topic is a corpus of 190-210 articles. 
    }
    \label{tbl:realistic_gpt4}
\end{figure}

Table~\ref{tbl:realistic_gpt4} shows the accuracy of our method in this evaluation under two different base models. Each topic's accuracy is evaluated over a randomly chosen test set containing $\%20$ of the articles and their simulated edited versions. We use the other $\%80$ as a training set to evaluate the per-length survival function $\bar{F}_{\Pglm;\Plm}$ and to calibrate the threshold of $\HC$ that maximizes the accuracy. 
%The results Figure~\ref{fig:realistic_gpt4} suggest that detecting the presence of edits in this dataset is generally more challenging than in the case reported in Table~\ref{tbl:results}, since the 
Standard analysis of the variance in the data shows that the performance significantly depends on the topic (in all $2 \times 3$ detection model and edit ratio combinations, the F-test's P-value is smaller than $10^{-7}$).

\subsection{Comparing to other approaches}
We compared the performance of our method to a classifier that only considers the minimal P-value $p_{(1)}$ in \eqref{eq:P-value} as in Bonferroni-type inference. Additionally, we considered a classification approach that may seen as standard in this challenge \citep{openAIdetector}: embedding articles using an LLM and using the embedded representation as features for a trained classifier such as in logistic regression. We also tried several trained classification methods when using word frequencies as features, but these methods were completely ineffective. We used the dataset described in Section~\ref{sec:realistic_edits}, with accuracy averaged over 10 splits in a cross-validation fashion: each split is used as a test set and the rest as a train set. Note that the training procedure for HC and minimal P-value ($\minP$) involves the characterization of the null LPPT distribution $G_0$ and estimating $\HC_{1-0.05}$. 

\begin{figure}
    \centering
    \begin{tikzpicture}
        \node (fig) at (0,0) {\includegraphics[scale=.4]{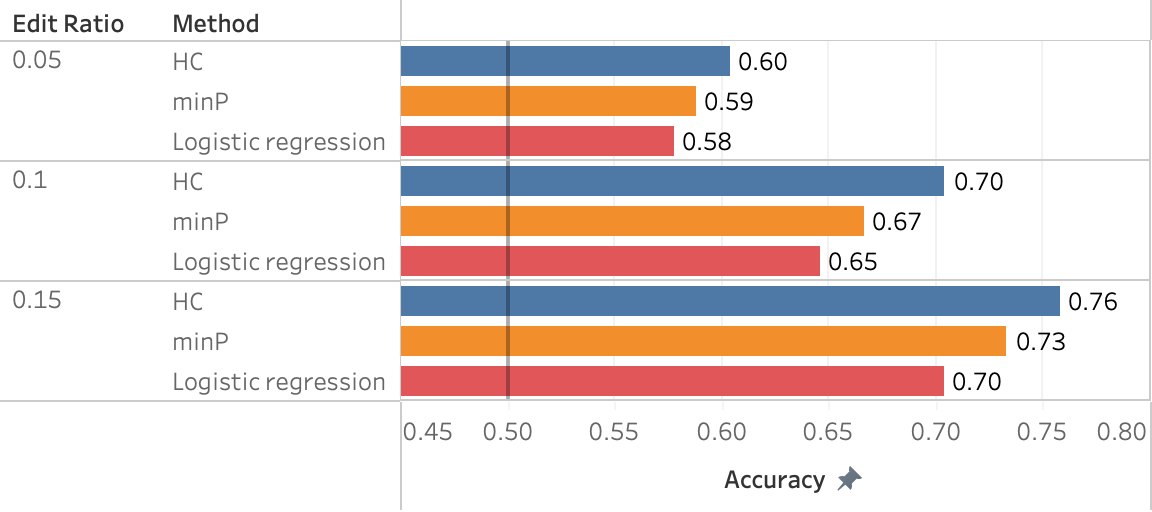}};
        \node[fill=white, below of=fig,node distance=2.3cm, xshift=3.7cm, yshift=0,minimum width=2cm,minimum height=0.5cm] { };
        %\node[below of=fig, node distance=2.4cm, xshift=-4.6cm] {Accuracy};
    \end{tikzpicture}
    \caption{Accuracy in detecting the presence of actual Wikipedia sentences planted within Wikipedia-style GPT3.5-turbo articles. Our method (HC) with detection model Phi2, inference based on the minimal P-value ($\minP$), and a logistic regression classifier with features obtained via document embedding using OpenAI's \texttt{text-embedding-3-small}. We used 10 train/test splits and reported the accuracy averaged over all splits. The standard error in all configurations is smaller than 0.02.  
 \label{fig:competing}}
\end{figure}

We report on the results of such comparison in Figure~\ref{fig:competing}. This figure shows that our HC-based approach attains the best accuracy in all configurations, significantly surpassing the trained classifier.

%We note that the better power of HC compared to inference based on the minimal P-value agrees with existing literature on the topic \citep{donoho2004higher,arias2015sparse,kipnis2021logchisquared}.

% \begin{table}[H]
% \begin{center}
% \input{table_competitive}
% \end{center}
%     \caption{
%     Our method versus a competing supervised method. The competing method is based on a linear classifier applied to features obtained by embedding articles with GPT4. 
%     \label{tbl:competing}}
% \end{table}

\begin{table}[H]
\begin{center}
\begin{tabular}{|lcc||ll|ll|}
\toprule
\multirow{2}{*}{\thead{article \\ title}} & 
\multirow{2}{*}{
\rotatebox[origin=c]{90}{length}} 
 & \multirow{2}{*}{\rotatebox[origin=c]{90}{edit ratio}} & 
 \multicolumn{2}{c|}{detecting using GPT2} & \multicolumn{2}{c|}{detecting using Phi2} \\
 \cmidrule(lr){4-5}
\cmidrule(lr){6-7}
& & & \multicolumn{2}{c|}{\thead{$\HC$ (P-value)}}
& \multicolumn{2}{c|}{\thead{$\HC$ (P-value)}} 
\\
\cmidrule(lr){4-5}
\cmidrule(lr){6-7}
& & & \thead{not edited} & \thead{edited} & \thead{not edited} & \thead{edited}\\
\midrule
Camus & 70 & 0.17 & 1.44 (0.2705) & {\color{blue} {3.45}} (\textbf{0.0003}) & {\color{red} {2.18}} (\textbf{0.0449}) & {\color{blue} {3.62}} (\textbf{0.0002}) \\
Kafka & 61 & 0.21 & 1.27 (0.365) & 1.75 (0.1404) & {\color{red} {2.38}} (\textbf{0.0237}) & {\color{blue} {3.0}} (\textbf{0.0021}) \\
Marquez & 53 & 0.19 & -1.4 (0.996) & -0.17 (0.9618) & 0.57 (0.7943) & 1.72 (0.1467) \\
Chekhov & 61 & 0.10 & 1.01 (0.5443) & 1.11 (0.473) & 1.32 (0.3376) & 2.03 (0.0694) \\
Morrison & 59 & 0.22 & -0.9 (0.9951) & 0.18 (0.9159) & -0.64 (0.9895) & 1.48 (0.2495) \\
Hesse & 66 & 0.20 & 0.83 (0.6677) & 2.04 (0.0689) & 1.18 (0.4293) & {\color{blue} {2.67}} (\textbf{0.01}) \\
Austen & 60 & 0.14 & 0.76 (0.7095) & {\color{blue} {2.44}} (\textbf{0.0196}) & 0.78 (0.6979) & {\color{blue} {2.96}} (\textbf{0.0024}) \\
Simenon & 67 & 0.18 & 1.38 (0.3052) & {\color{blue} {2.58}} (\textbf{0.0134}) & {\color{red} {2.78}} (\textbf{0.0065}) & {\color{blue} {3.73}} (\textbf{0.0001}) \\
Tolstoy & 64 & 0.24 & -0.1 (0.9566) & 1.55 (0.2182) & 0.65 (0.7646) & {\color{blue} {2.22}} (\textbf{0.0397}) \\
Rice & 51 & 0.18 & 0.63 (0.7652) & 1.51 (0.2283) & 1.11 (0.4555) & {\color{blue} {2.4}} (\textbf{0.0214}) \\
Rowling & 58 & 0.27 & 0.95 (0.5701) & 1.72 (0.1481) & 1.37 (0.3015) & {\color{blue} {3.04}} (\textbf{0.0015}) \\
Andersen & 80 & 0.29 & 0.87 (0.647) & {\color{blue} {3.97}} (\textbf{0.0001}) & 1.38 (0.313) & {\color{blue} {4.47}} (\textbf{0.0001}) \\
Mishima & 55 & 0.21 & 0.84 (0.6447) & {\color{blue} {2.65}} (\textbf{0.0093}) & 1.69 (0.153) & {\color{blue} {4.09}} (\textbf{0.0001}) \\
Verne & 64 & 0.23 & -0.07 (0.9535) & 0.59 (0.7961) & 0.53 (0.8153) & 1.26 (0.3795) \\
Dickinson & 68 & 0.26 & 1.65 (0.1769) & {\color{blue} {3.9}} (\textbf{0.0001}) & 1.32 (0.3357) & {\color{blue} {4.45}} (\textbf{0.0001}) \\
Po & 51 & 0.21 & -0.21 (0.9652) & 0.59 (0.7855) & 0.78 (0.6838) & {\color{blue} {2.14}} (\textbf{0.0485}) \\
Christie & 78 & 0.20 & -0.23 (0.9725) & 0.6 (0.8032) & 0.23 (0.9111) & 1.75 (0.1484) \\
Dahl & 58 & 0.15 & 1.02 (0.5407) & 1.81 (0.1237) & 0.91 (0.6171) & 1.1 (0.476) \\
\bottomrule
\end{tabular}
\end{center}
    \caption{
Detecting a few human edits in Wikipedia-style biographies written by ChatGPT with a perplexity detector based on GPT2. The table shows HC of \eqref{eq:HC_def} and the P-value of the HC test based on simulated values as in Figure~\ref{fig:critical_values}. Also shown are the fraction of edited sentences in every document and its length $n$ in sentences. Values significant at level $\alpha=0.05$ are in color: blue for a true positive and red for a false positive. In all cases, Bonferroni's correction $p_{(1)} \times n$ was insignificant, mainly because the number of sentences in the empirical P-value calculations of the LPPT test \eqref{eq:P-value_w_length} is small. 
    \label{tbl:results_real}}
\end{table}

% \begin{table}[H]
% \begin{center}
% \input{table_results_real_bio_phi}
% \end{center}
%     \caption{
%     Detecting a few human edits in Wikipedia-style biographies written by ChatGPT (GPT3.5) with a perplexity detector based on the model GPT2xl. 
%     The table shows HC of \eqref{eq:HC_def} and the P-value of the HC test based on simulated values as in Figure~\ref{fig:critical_values}. HC values significant at level $\alpha=0.05$ are in blue (indicate true positives). HC values of non-edited documents significant at level $\alpha=0.05$ are in red (indicate false positives). Also shown are the fraction of edited sentences in every document, its length $n$ in sentences, and the minimal P-value times $n$ associated with Bonferroni correction.  
%     \label{tbl:results_real}}
% \end{table}

\subsection{Manually edited articles}
In Table~\ref{tbl:results_real} we report on the results of applying our method to \nDocs~ articles that were created via the following process: we used the GLM GPT3.5-turbo\footnote{Via the ChatGPT web interface \url{https://chat.openai.com/}.} to write a Wikipedia-style biography of one of the authors in the list according to a prescribed structure involving 4 sections: Early Life, Adulthood, Contributions and Achievements, and Legacy; the GLM wrote each section in response to a separate prompt, similarly to the process illustrated in Figure~\ref{fig:generation}. We used the text written by the GLM and the prescribed subtitles to form a coherent article which we denote as the pre-edited GLM article. Next, we asked a human editor to modify this article by adding, rephrasing, or removing entire sentences. We applied our method to both the edited and non-edited articles. We used sentences from additional articles created similarly to characterize $\bar{F}_{\Pglm,\Plm}$ and evaluate the P-values in \eqref{eq:P-value_w_length}.

Table~\ref{tbl:results_real} shows that all edited articles have larger HC values than their non-edited version. We also report on the P-values associated with these HC values under the null of uniformly distributed P-values obtained via simulations. We also evaluated $\text{(num. of sentences)} \times \min p_i$ in each article, 
which is associated with the significance of the Bonferroni correction applied to the body of sentence P-values. These values turned out to be above $0.05$ in all articles. This peculiar behavior of the Bonferroni correction is because the minimal LPPT P-values in our process are approximately $1/(1+n_\ell)$, where $n_\ell$ is the number of GLM sentences in our calibration data of length $\ell$. However, in most lengths, $n_\ell$ is relatively small at the order of $10^2$. In a future analysis, we may make Bonferroni's correction more useful by increasing our calibration set or by fitting a curve to the tail of the survival function in Figure~\ref{fig:refinements} and verifying the goodness of this fit. 
% Table~\ref{tbl:results_real} summarizes the values reported in Table~\ref{tbl:summary_of_real_results}. %(false positives after Bonferroni's correction are not indicated in Table~\ref{tbl:results_real}.)
% \begin{table}[H]
% \begin{center}
% \begin{tabular}{|c|c|c|c|c|}
% \toprule
%     detection model & \multicolumn{2}{c|}{GPT2} & \multicolumn{2}{c|}{Phi2} \\
%   & $\HC$ & Bonferroni & $\HC$ & Bonferroni \\
%      \midrule
%      True Positive Rate  & 7/18  & 7/18 & 11/18 & 8/18 \\
%      False Positives Rate & 0/18 & 1/18 & 1/18 & 2/18 \\
%      \bottomrule
% \end{tabular}
% \end{center}
% \caption{Summary of the results of Table~\ref{tbl:results_real} (False positives after Bonferroni's correction  are not indicated in  Table~\ref{tbl:results_real}.).}
% \label{tbl:summary_of_real_results}
% \end{table}
We emphasize that the discrimination reflected in Table~\ref{tbl:results_real} is without calibrating the threshold of $\HC$ for separating the class of edited from non-edited articles. Such calibration is expected to increase the accuracy of the procedure.

%We note that although this observation supports the effectiveness of the method, a much larger sample of articles is needed to verify that a test based on these P-values is indeed of the size prescribed due to possible violations of the independent sentences model \eqref{eq:global_null}. 

%In Table~\ref{tbl:HCT} we show the sentences included in the set $I^*$ for the article \texttt{Welsh Corgi}. 

\section{Information-Theoretic Analysis
\label{sec:IT}}

We now analyze our method under a theoretical framework of text editing and discuss some factors affecting its success.

\subsection{Optimality of the Higher Criticism test}
A simple mixture model for the generation of an edited document proposes that most sentences are written independently by a GLM $\Pglm$, except perhaps a few sentences that are generated by a different mechanism associated with the editor that we denote here by $\Palt$. Importantly, we do not know in advance which sentences were written by each model. Let $\epsilon$ denote the expected fraction of $G_1$ sentences, and let $L_j$ be the distribution of $\logl(S;\Plm)$ under $S\sim G_j$, for $j\in \{0,1\}$. The setting described above induces a mixture model for the log-perplexity
\begin{align}
\begin{split}
H_0 \,&:\, \logl(S_i; \Plm) \simiid L_0,\qquad i=1,\ldots,n,  \\
H_1\,& :\, \logl(S_i; \Plm) \simiid (1-\epsilon) \cdot  L_0 + \epsilon\cdot  L_1,\quad i=1,\ldots,n.
\end{split}
\label{eq:mixture_lloss}
\end{align}
Likewise, we have a mixture model for the P-values in \eqref{eq:P-value}:
\begin{subequations}
\label{eq:mixture_pvalues}
\begin{align}
H_0\,& :\,p_i \simiid \Unif(0,1),\qquad i=1,\ldots,n, \label{eq:H0} \\
H_1\,&:\, p_i \simiid (1-\epsilon) \cdot \Unif(0,1) + \epsilon \cdot Q_i,\quad i=1,\ldots,n, \label{eq:H1}
\end{align}
\end{subequations}
where here $Q_i$ is a sub-uniform distribution that describes the non-null behavior of the P-values \eqref{eq:P-value}. The optimality of HC for mixture models of the form \eqref{eq:mixture_lloss} and \eqref{eq:mixture_pvalues} have been studied in several contexts  \citep{donoho2004higher,jin2003detecting,hall2008properties,cai2014optimal,arias2015sparse,mukherjee2015hypothesis,jin2016rare,DonohoKipnis2022,kipnis2021logchisquared}. In particular, when the mixture parameter is calibrated to $n$ as $\epsilon = n^{-\beta}$, for some $\beta \in (1/2,1)$, and the effect size in $Q_i$ is moderately large, a test based on HC of $p_1,\ldots,p_n$ attains the information-theoretic limit of detection in \eqref{eq:mixture_pvalues} when $n \to \infty$. Namely, in a configuration of the calibrating parameters in which there exists a test of asymptotically non-trivial power, there exists a test based on HC that is asymptotically powerful in the sense that its power tends to one while its size tends to zero. 

The works of \citep{delaigle2009higher,hall2008properties,hall2010innovated} extended the optimality properties of HC to some situations of dependent individual effects, unlike the model \eqref{eq:mixture_pvalues}. One relevant conclusion from these works is that HC is relatively unaffected when the P-values experience a form of short-term dependency as expected among sentences.

\subsection{Optimality properties of the perplexity test}
The justification for using the LPPT test of \eqref{eq:P-value} is primarily its empirical success in separating GLM from non-GLM sentences shown in Figure~\ref{fig:individual_chunks}. In what follows, we analyze the power of this test beyond this empirical observation. 

\subsubsection{Language model as an information source and the asymptotic perplexity}
Let $\Pa$ be a language model. Sampling a sentence $t_{1:n} = (t_1,\ldots,t_n)$ form $\Pa$ is achieved by conditioning current token probability by previous tokens and an initial context. Namely, 
    \begin{align}
        \label{eq:sampling_from_lm}
    t_i \sim \Pa(\cdot |t_0, t_{1:i-1})\Pa(t_0),\quad i=1,\ldots,n,
    \end{align}
for some initial state $t_0$ that can represent some initial context like the text's topic or a null value. We view $\Pa$ as an information source in the sense that it defines a stationary distribution over sequences of tokens from a finite alphabet \citep{shannon1948mathematical}. When $\Pa$ is ergodic, the Shannon-McMillan-Brieman theorem says that the entropy rate $\Hcal(\Plm)$ is well-defined by the limit 
\begin{align}
    \label{eq:logloss_lim}
\Hcal(\Pa) = \lim_{n\to \infty} \logl(t_{1:n}; \Pa),
\end{align}
which is independent of $t_0$ \citep{algoet1988sandwich}. In the absence of ergodicity, the limit \eqref{eq:logloss_lim} may still exist but it generally depends on the initial state \citep{katok1977theory}. We note that Shannon's source coding theory proposes an alternative definition of the entropy rate: the minimum number of expected bits per token needed to represent $t_{1:n}$ as $n \to \infty$ \citep{cover1978convergent}.
%This operational definition is valid even if $\Pa$ is not ergodic, provided the expectation is also over the initial state $t_0$ \citep{gray2011entropy}. 

More generally, suppose that we evaluate the LPPT with respect to another stationary probability law $\Pb$ defined over the same alphabet as $\Pa$. Under some conditions on the laws $(\Pa, \Pb)$, the limit of $\logl(t_{1:n}; \Plm)$ as $n\to \infty$ exists almost surely and obeys
\begin{align}
\label{eq:cross_entropy_asymp}
\lim_{n\to \infty} \logl(t_{1:n}; \Pb) = \Hcal(\Pb; \Pa) = \Hcal(\Pa) + \Dkl(\Pa || \Pb),
\end{align}
where $\Dkl(\Pa || \Pb)$ is the relative entropy rate of $\Pb$ to $\Pa$ \citep[Ch. 7]{gray2011entropy}. The term $\Hcal(\Pb; \Pa)$ is
denoted as the crossentropy rate of $\Pb$ under the law $\Pa$. 
%Similarly to the entropy rate, the crossentropy rate also admits an operational definition as the expected number of bits per token used in representing a long sequence from $\Pa$ using a binary code that, if applied to long sequences from $\Pb$, would converge to $\Hcal(\Pa)$ \citep{cybenko1998mathematics,gray2003mismatch}. A practical lossy compression method that presumably asymptotically attains an expected code length $\Hcal(\Pb;\Pa)$ is an arithmetic encoder applied to every token with interval partitions based on $\Pb$. An extension of this compression method that was implemented in \citep{izacardlossless} is known to attain state-of-the-art results in text compressing large texts \citep{mahoney2011large}.
Relation \eqref{eq:cross_entropy_asymp} appears to provide an interesting insight about the LM $\Plm^*$ that maximizes the power of the LPPT for testing $H_{0,S}$ of \eqref{eq:null_sentence} versus a simple alternative 
\begin{align}
    \label{eq:H1S}
    H_{1,S} \, :\, S \sim \Palt 
\end{align}
for some information source $\Palt$ that represents the effect of editing the sentence $S$. 
%\subsubsection{Desirable properties of $\Plm$}
Suppose that $\Pglm$ and $\Palt$ are fixed, i.e. determined by the problem's nature. We seek a base model $\Plm$ for the perplexity detector that maximizes the power of the perplexity test in \eqref{eq:P-value}. The companion article \citep{kipnis2024Iprojection} shows that under some reasonable assumptions, the ideal base model $\Plm$ maximizes 
\begin{align}
\label{eq:delta_def}
    \Delta(\Palt, \Pglm; \Plm) := \Dkl(\Palt || \Plm) - \Dkl(\Pglm || \Plm),
\end{align}
where $\Dkl$ indicates the relative entropy rate of information sources \citep{gray2011entropy}. We discuss possible implications of \eqref{eq:delta_def} in Section~\ref{sec:optimal_p} below.

\section{Avenues for Future Research}
\label{sec:challenges}

\subsection{Incorporating context}
\label{sec:context}
Typically, a sentence written by a GLM depends on the previous sentence or another context affecting the GLM's state. The effect of the context on the perplexity may be significant due to a potential lack of ergodicity in actual writings or slow convergence of the LPPT to its limiting value if such convergence occurs. For this reason, it appears that incorporating a context in the LPPT evaluations 
may increase the power of the perplexity detector over individual sentences. Denote the LPPT of a sentence $S = (t_1,\ldots,t_{|S|})$ and context $C$ as
\begin{align}
    \label{eq:lloss_ctx}
    \logl(S;C,\Plm) := -\frac{1}{|S|} \sum_{i=1}^{|S|} \log \Plm(t_i | t_{1:i-1}, C). 
\end{align}
The context $C$ is usually a sequence of tokens, e.g., the sentence preceding $S$, although it may also take other forms such as the activations of the attention mechanism in transformers-based language models \citep[Chapter 11]{jurafsky2023}. If the policy determining $C$ is also stationary (e.g., the preceding sentence policy), we can extend much of the analysis in Section~\ref{sec:IT} to use \eqref{eq:lloss_ctx} instead of \eqref{eq:LL_def}.

\subsection{Maximizing the power of the perplexity detector
\label{sec:optimal_p}
}
Our analysis in Section~\ref{sec:IT} shows that the power of the log-perplexity detector is proportional to the difference in relative entropy $\Delta(\Palt, \Pglm; \Plm)$ of \eqref{eq:delta_def}. The information projection principle \citep{csiszar2003information,cover2006elements} implies an interesting direction in searching for $\Plm$ that maximizes this difference. Informally, suppose that we search for a maximizer $\Plm^*$ within a convex set $\mathcal P$ of available LMs. We assume that
\begin{align}
\label{eq:mathcalp}
\Dkl(\Palt||\Pglm) \leq \Dkl(\Palt||\Plm),\qquad \Plm \in \mathcal P.
\end{align}
Namely, the candidate GLM is closer in relative entropy to the alternative model than any of the LMs in our search space $\mathcal P$. This situation is justified, e.g., because $\Pglm$ is optimized to mimic the behavior of $\Palt$ which represents human writing. This optimization is achieved primarily via log-perplexity minimization which asymptotically translates to relative entropy minimization \citep{jurafsky2023}.
Now, by the information projection principle \citep[Ch. 11]{cover2006elements}
\[
\Dkl(\Palt||\Plm) \geq \Dkl(\Palt||\Pglm)  + \Dkl(\Pglm||\Plm),\qquad \Plm \in \mathcal P,
\]
and thus
\begin{align*}
    \Delta(\Palt, \Pglm; \Plm) \geq \Dkl(\Palt;\Pglm), \qquad \Plm \in \mathcal P.
\end{align*}
The last inequality is attained with equality when $\Plm = \Pglm$, implying that this choice of $\Plm$ is the worst choice over models with the property \eqref{eq:mathcalp}. Specifically, a better choice of $\Plm$ should also consider the relative entropy to the alternative model $\Palt$. The characterization of such an alternative model in applications appears to be challenging, although the relative entropy can be approximately evaluated using standard methods, e.g. via the excessive binary code length in lossless compression \citep{cybenko1998mathematics,gray2003mismatch}.   

%(although this case may not be realizable because we may not evaluate the token probabilities), under which case $\Delta(\Palt, \Pglm; \Plm) = \Dkl(\Palt;\Pglm)$. 

%$\Pglm$ on the convergence of $\logl(S;\Plm)$ to $\Hcal(\Plm;\Pglm)$. A related study is \citep{basu2020mirostat}, which analyses the Effect of different methods on sampling from a LM on the log-perplexity. 
%In principle, one can rank all approximating models $\Plm$ to the GLM by their relative entropy $\Dkl(\Plm, \Pglm)$. Indeed, assuming that the LPPT is a good estimator to $\Hcal(\Plm; \Pglm)$ due to the convergence in \eqref{eq:cross_entropy_asymp}, one may estimate $\Dkl(\Plm, \Pglm)$ by subtracting an estimate of $\Hcal(\Pglm)$ from $\Hcal(\Plm; \Pglm)$. An estimate of $\Hcal(\Pglm)$ may be obtained using standard entropy estimation methods via a long enough text written by the GLM \citep{wyner1989some,kontoyiannis1998nonparametric}. 

\subsection{Assessing the minimal number of edits for detection}

The connection between the problems of detecting edits and rare signals this paper promotes suggests the possibility of estimating the minimal number of edits one must make so that detecting their global presence is possible.

A great deal of literature discussed the tradeoff between the rarity and the magnitude of individual non-null effects in sparse signal detection \citep{Ingster1993,jin2003detecting,donoho2004higher,delaigle2009higher,cai2014optimal,mukherjee2015hypothesis,DonohoKipnis2022,kipnis2021logchisquared}. In particular, suppose that we have $n$ asymptotically normal and independent tests, in which non-null effects are on the moderate deviation scale. On the P-value scale, this can be written as
\begin{align}
    \label{eq:P-value_moderatae}
p_i \overset{D}{\approx} \bar{\Phi}\left( \mu_n + \sigma Z  \right),\qquad Z \sim \Ncal(0,1), 
\end{align}
where $\mu_n = \sqrt{2 r \log(n)}$ for some signal intensity parameter $r >0$, $\sigma>0$, and $\overset{D}{\approx}$ is a form of asymptotic equivalence in distribution that is described in \citep{kipnis2021logchisquared}. Additionally, assume that the proportion of non-null effects is $\epsilon = n^{-\beta}$, $\beta \in (0,1)$. The asymptotic power of detecting the global significance of the body of P-values experiences a phase transition as $n \to \infty$, described by the curve
\begin{align}
\rho^*(\beta,\sigma) & :=
\begin{cases}
     (2 - \sigma^2)(\beta - 1/2) & \frac{1}{2} < \beta < 1- \frac{\sigma^2}{4}, \quad 0<\sigma^2<2,\\
    \left(1-\sigma\sqrt{1 -\beta }\right)^2 &  1- \frac{\sigma^2}{4} \leq  \beta < 1, \quad 0<\sigma^2<2,\\
     0 & 0 < \beta <  1-\frac{1}{\sigma^2}, \quad \sigma^2 \geq 2,\\
    \left(1-\sigma\sqrt{1 -\beta }\right)^2 &  1-\frac{1}{\sigma^2} \leq  \beta < 1, \quad \sigma^2 \geq 2.
    \end{cases}
    \label{eq:rho}
\end{align}
Namely, if $r < \rho(\beta;\sigma)$ any test distinguishing the null hypothesis of uniformly distributed P-values from a situation that $\approx n^{1-\beta}$ of them obey \eqref{eq:P-value_moderatae} has asymptotically trivial power. If $r > \rho(\beta;\sigma)$, some tests, such as HC of \eqref{eq:HC_def}, are asymptotically fully powered in the sense that there exists a sequence of thresholds under which the sum of Type-I and Type-II errors goes to zero. Consequently, \eqref{eq:rho} describes the way the number of non-null effects $n \epsilon = n^{1-\beta}$ must scale with $n$ to guarantee a non-trivial power of HC or any other global testing procedure. For example, if $\sigma=1$ and $r \leq  1/4$, detection is possible when $\epsilon = \Omega \left( n^{1/2-r}\right)$.

This description provides an estimate to the number of edits necessary for reliably detecting the presence of any edits, provided one can estimate the signal strength parameters $r$ and $\sigma$ associated with two GLMs from the data. A standing challenge in this context is analyzing the usefulness of this estimate, e.g. by establishing the relevance of the model \eqref{eq:mixture_pvalues} with perturbations of the form \eqref{eq:P-value_moderatae} in practice and real data evaluations.

\bibliographystyle{alpha} 
\bibliography{amain_IEEE}

\subsection*{Acknowledgments}
 The author would like to thank David Donoho for helpful discussions regarding the problem formulation, Mira Kipnis for creating and editing ChatGPT articles, Boaz Zimbler for helping select the models for perplexity detection, and Adam Vinestock and Omer Wachman for studying the perplexity detector with different context policies.
% \appendix
% \input{appendix}

\end{document}